\documentstyle[12pt,aasms4,psfig]{article}

\slugcomment{to appear in the Astronomical Journal}  
 
\begin{document}
 
\title{Radio Observations of Infrared Luminous High Redshift QSOs}
 
\author{
C. L. Carilli$^{1}$,
F. Bertoldi$^2$,
A. Omont$^3$,
P. Cox$^4$,
R.G. McMahon$^5$,
K.G. Isaak$^6$
}          

\affil{$^{1}$National Radio Astronomy Observatory, P.O. Box O,
Socorro, NM 87801, USA, ccarilli@nrao.edu}      

\affil{$^{2}$Max-Planck-Institut f\"{u}r Radioastronomie,
Auf dem H\"ugel 69, D-53121 Bonn, Germany}

\affil{$^{3}$Institut d'Astrophysique de Paris, CNRS, 98bis boulevard
Arago, F-75014, Paris, France}

\affil{$^{4}$Institut d'Astrophysique Spatiale, Universite de Paris XI,
F-91405 Orsay, France} 

\affil{$^{5}$Institute of Astronomy, Madingley Road, Cambridge, CB3
0HA, UK}

\affil{$^{5}$Cavendish Laboratory, Madingley Road, Cambridge, CB3
0HE, UK}

\begin{abstract}

We present Very Large Array (VLA) observations at 1.4 GHz and 5 GHz  of a 
sample of 12 Quasi-stellar Objects (QSOs) at $z = \rm 3.99~to ~ 4.46$.
The sources were selected as the
brightest sources at 250 GHz from the recent survey  of
Omont et al.  (2001).   We detect seven sources  at 1.4 GHz with flux
densities, $\rm S_{1.4} \ge  50~ \mu Jy$.  These
centimeter (cm) wavelength observations
imply that the millimeter (mm)  emission 
is most likely thermal dust
emission. The radio-through-optical spectral
energy distributions for these sources are within the broad range
defined by lower redshift, lower optical luminosity QSOs. 
For two sources the
radio continuum luminosities and morphologies indicate steep spectrum,
radio loud emission from a jet-driven radio source.
For the remaining 10 sources the 1.4 GHz
flux densities, or limits, are consistent with those expected for
active star forming galaxies.  If the radio emission is 
powered by star formation in these systems, then the implied
star formation rates are of order 10$^3$ M$_\odot$ year$^{-1}$. 
We discuss the angular sizes and spatial distributions
of the radio emitting regions,
and we consider briefly these results in the context of co-eval 
black hole and stellar bulge formation in galaxies. 

\end{abstract}
 

\section {Introduction}

Recent study of the dynamics of gas and stars in the nuclear regions
of nearby galaxies has led to two remarkable discoveries: (i) the
overwhelming majority of spheroidal galaxies in the nearby universe
contain massive black holes ( Tanaka et al. 1995; Miyoshi et
al. 1996; Richstone et al. 1998; Ghez et al. 1998; 
van der Marel 1999;  Genzel et al.
2000; Kormendy and Ho 2000), and (ii) there is a clear correlation between
the black hole mass and the velocity dispersion of the stars in the
spheroid (Gebhardt et al. 2000, Ferrarese \& Merritt 2000). This
latter correlation suggests a `causal connection between the formation
and evolution of the black hole and the bulge' (Gebhardt et al. 2000),
meaning that black holes may be 
a fundamental component of galaxies, and not simply exotica giving
rise to AGN phenomenon (Richstone et al. 1998). A
possible explanation for this correlation is that massive black hole
and spheroid formation are co-eval, occurring during the collapse of a
galaxy from the primordial density field. Such a model is motivated by
the fact that the correlation of black mass with velocity dispersion
is significantly tighter than that with spheroid luminosity, implying
that systems with greater dissipation during collapse, ie. high
surface brightness, compact spheroids with high velocity dispersion,
form larger black holes.

Possible supporting evidence for co-eval starbursts and AGN at high
redshift is the detection of copious emission from warm dust 
in high redshift QSOs (McMahon et al. 1994;
Omont et al. 1996a; Andreani et al. 1999; Carilli et
al. 2000; Omont et al. 2001; Carilli et al. 2001a; Isaak et al. 2001).
Searches with mJy 
sensitivity at 250 GHz have yielded detections of 41 QSOs at $z = 
\rm 3.7 ~ to ~5.0$, out of a total of 147 sources observed. The implied
Far Infrared (FIR) luminosities are of order $10^{13}$ L$_\odot$ with
implied dust masses of a $\rm few \times 10^8$ M$_\odot$.  Follow-up
observations of three 
of these dust-emitting QSOs revealed CO emission as well, with implied
molecular gas masses of a $\rm few \times 10^{10}$ M$_\odot$ (Guilloteau et
al. 1997, 1999; Ohta et al. 1996; Omont et al. 1996b; Carilli, Menten,
\& Yun 1999). Omont et al. (2001) argue that the 
dominant dust heating mechanism may be
star formation based on the two simple points: (i) 
star formation is inevitable given the large dust and gas masses, 
and (ii) star formation is required to produce the large dust masses. 
The implied star formation rates are so extreme,
$\ge$ 10$^3$ M$_\odot$ year$^{-1}$, that a significant fraction of the
stars in the QSO host galaxy can be formed on galaxian dynamical
timescales. Alternatively, Sanders et al. (1989) argue that radiation
from the AGN is the dominant dust heating mechanism, since in most cases it
requires the absorption of only a small fraction ($\le 20\%$) of the
AGN UV luminosity.  

A possible means of addressing the question of dust heating by star
formation vs. AGN radiation is through radio observations.  Star
forming galaxies at low redshift follow a very tight, linear
correlation between radio continuum and FIR luminosity (Condon 1992).
This correlation holds over four orders of magnitude in luminosity
with only a factor two scatter around linearity for galaxy samples
selected in the optical, IR, and radio. A general correlation between
synchrotron radiation at cm wavelengths and thermal dust emission at
IR wavelengths is expected, since both relate to massive star
formation, however the tightness and linearity of the correlation
remain puzzling. If this correlation
holds to high redshift (Carilli and Yun 1999, 2000; Yun et al. 2000),
then comparing cm and mm continuum luminosities is a possible
diagnostic of star formation.

In this paper we present VLA observations at 1.4 GHz and 5 GHz of a
sample of high redshift QSOs from the Digitized Palomar Sky Survey
(PSS; Kennefick et al. 1995 a, b; Djorgovski et al. 1999)
and the Automatic Plate Measurement survey
(APM; Storrie-Lombardi et al. 1996).  The sources
were selected as the brightest sources at 250 GHz from the recent
survey of Omont et al. (2001). These radio observations are more
sensitive than previous radio studies of  high redshift QSOs
(Schmidt et al. 1995b; Stern et al. 2000), and in
particular, probe to levels that could indicate star formation, as
dictated by the radio-to-FIR correlation. We also consider the radio
continuum morphologies and spectral indices in this regard.  We assume
$\rm H_o = 50$ km s$^{-1}$ Mpc$^{-1}$, $\rm q_o = 0.5$, and we define
the spectral index, $\alpha$, as a function of frequency, $\nu$, as:
$\rm S_\nu \propto \nu^\alpha$.
 
\section{Observations}

Our sample consisted of the 12 strongest sources detected at
250 GHz by Omont et al.  (2001). The sources are listed in Table 1,
and their positions are given in Omont et al.  (2001). 
For most of the sources the 1.4 GHz VLA observations were made in the
A (30 km) configuration 
in December 2000 using a total bandwidth of 100 MHz with two
orthogonal polarizations.  Each source was observed for 2
hours. For three of the sources (B1117--1329, B1144--0723,
J1646+5514)  observations were made in the B (10 km)
configuration in April 2001. 

Standard phase and amplitude calibration was applied,
as well as self-calibration using background sources in the telescope
beam. The absolute flux density scale was set using observations of 3C
286.  The final images were generated using the wide field imaging and
deconvolution capabilities of the AIPS task IMAGR.  The theoretical
rms noise ($\sigma$) value is between 15 and 25 $\mu$Jy beam$^{-1}$,
depending on the range of telescope elevations over which the source
was observed, and for most of the sources the measured noise values
are in this range. For two sources (J1253-0228 and B1144--0723) the 
noise level is
significantly higher due to side-lobe confusion by bright sources 
within the primary beam of the telescope.
The Gaussian restoring CLEAN beam Full Width at Half Maximum (FWHM)
was typically 1.5$''$ 
for the A configuration observations, and
$8''\times 4''$ (major axis  position angle = $-40^o$)
for the B configuration observations of B1117--1329 and
B1144--0723. The source J1646+5514 was observed at low elevation, 
leading to a highly elongated beam (FWHM = $21''\times4''$, with a
major axis position angle of 52$^o$), and a higher noise level
due to ground pick-up.
One source, J1057+4555, was detected as part of the FIRST survey
(Becker, White, \& Helfand 1995; Stern et al. 2000).  We re-observed
this source, and obtained 
the same flux density to within 6$\%$. 

We observed the four brightest sources detected at 1.4 GHz (J1048+4407,
J1057+4555, J1347+4956, J2322+1944) with the VLA at 5 GHz in January 2001
in the A configuration.  Each source was observed for 1 hour,
resulting in rms noise values between 25 and 35 $\mu$Jy.  Standard
calibration and imaging was performed, using 3C 286 as the absolute
flux calibrator. The FWHM of the Gaussian restoring beams were about
0.5$''$.  The 5 GHz images were then convolved to the resolution of
the 1.4 GHz images to determine spectral indices.

\section{Results}

Table 1 lists the results for the 12 sources observed with the VLA.
Column 3 gives the 250 GHz flux densities, column 4 gives the
integrated 1.4 GHz flux densities, and Column 5 gives the 350 GHz flux
densities  from Isaak  et al. (2001). 
The M$_{\rm B}$ in Column 6 are values are from
Isaak et al. (2001) and McMahon et al. (1999). 

The 1.4 GHz images of these sources
are shown in Figure 1.  The positional uncertainty 
for the radio observations is
given by: $\rm \sigma_\theta \sim {{FWHM}\over{SNR}}$ (Fomalont 1998),
where FWHM corresponds to that of the 
Gaussian restoring beam, and SNR = Signal-to-Noise ratio of the
detection. For a $3\sigma$ detection this 
corresponds to $0.5''$ for most of our sources, except
the three sources that were observed in the B array, for which 
the positional uncertainty is increased by about a factor three.
To this must be
added (in quadrature)
the  positional uncertainty in the optical, for which an rms
accuracy of about 0.5$''$ has been estimated (McMahon priv comm),
and the uncertainty in the relationship between 
radio and optical reference frames, which is about
$0.2''$ (Stone 1994). Overall, we consider a  reasonable detection as
a source with a flux density $\ge 3\sigma$ within 0.7$''$ of the
optical QSO position. These sources are marked with a `d' in
Column 4 of Table 1. Faint radio source counts at 1.4 GHz imply that
the chance probability of detecting a 50$\mu$Jy source within
0.7$''$ of a given QSO position is $3\times 10^{-4}$ (Richards 2000).

\subsection{Results on Individual Sources}

{\bf PSS J0808+5215} A radio continuum source with S$_{1.4}$ = 
58$\pm$18 $\mu$Jy is detected at  (J2000) 
$\rm 08^h 08^m 49.43^s$ $+52^o 15' 14.9''$, within  0.3$''$
of the optical QSO position. The optical spectrum of this source
is not yet published.

{\bf PSS J1048+4407} This source is clearly detected at 1.4 GHz. The
position of the radio continuum peak is (J2000) 
$\rm 10^h 48^m 46.63^s$ $+44^o 07' 10.8''$.
Gaussian fitting to the radio emission sets an upper limit to
the source size of 1.1$''$ at 1.4 GHz.  The 5 GHz image of this source
is shown in Figure 2. The source is also detected at 5 GHz, with a
total flux density of 179$\pm$37 $\mu$Jy, as determined from images
convolved to the resolution at 1.4 GHz, and the position of the 5 GHz
source at this resolution is within 0.2$''$ of that determined at 1.4
GHz.  The spectral index between 1.4 and 5 GHz is --0.70.  The
$0.5''$ resolution 5 GHz image suggests a resolved source
with a peak surface brightness of 105$\mu$Jy beam$^{-1}$. Interestingly,
this peak  is  located 0.4$''$ west of the optical
QSO position.  We feel it likely that this source is a radio jet
source, and that the $0.5''$ resolution 5 GHz image reveals a radio hot
spot, however, more sensitive, high resolution images are clearly
required. The optical spectrum shows that source this is
a  broad absorption line quasar (Peroux et al. 2001)

{\bf PSS J1057+4555} This source is clearly detected at 1.4 GHz, and
the radio morphology reveals a double lobed source, roughly symmetric
about the optical QSO position, with a total extent of about
4$''$. The $0.5''$ resolution 5 GHz image of this source is shown in
Figure 2. No flat spectrum radio nucleus is detected, and only faint,
diffuse radio emission is detected in the vicinity of the brighter
(southwestern) radio lobe. Convolving to the 1.4 GHz resolution
results in a spectral index for the southern lobe of --0.92, and a
(2$\sigma$) upper limit to the spectral index of the northern lobe of
--0.5. The optical spectrum published by 
Peroux et al. (2001) shows this to be a normal emission line quasar.

{\bf APM BR B1117--1329} This source is not detected in the B
configuration  observations with a 
3$\sigma$ upper limit of S$_{1.4} < 0.1~\mu$Jy.
The optical spectrum shows this source to be a broad
absorption line quasar (Storrie-Lombardi et al. 1996)

{\bf APM BR  B1144--0723} This source is not detected in the B
configuration observations with a 
3$\sigma$ upper limit of S$_{1.4} < 0.12~\mu$Jy.
The optical spectrum shows this source to be a broad
absorption line quasar (Storrie-Lombardi et al. 1996).

{\bf PSS J1248+3110} A source with S$_{1.4}$ = 71$\pm 15$ $\mu$Jy is
detected at (J2000) $\rm 12^h 48^m 20.2^s$, $+31^o 10' 43.37''$, 
0.8$''$ south of the optical QSO position. Based on the
faint radio source counts in Richards (2000), 
the probability of detecting a 71$\mu$Jy source at 1.4 GHz within
0.8$''$ radius of a given QSO position 
by chance is $4\times10^{-4}$. Also, given that this is 
close to a 5$\sigma$ detection, the probability that it is simply
a noise feature is comparably small. We feel it likely
that this radio source is real, and somehow associated with 
the QSO.  Taking into consideration the positional uncertainties
in the radio and optical (see section 3), 
the positional offset between the radio source and 
the optical QSO is significant at only  the 
$1.3\sigma_\theta$ level. In the analysis below, we consider the
radio emission to be from the QSO itself, although it
remains possible that  we are seeing 
radio emission from a companion galaxy. 
Gaussian fitting to the radio emission sets an upper limit to
the source size of 1.8$''$ at 1.4 GHz. 
The optical spectrum of this source is not yet published.

{\bf PSS J1253--0228} This source was not detected at 1.4 GHz, although the
noise in this image is a factor 2.5 higher than for most of the other
sources, such that the 3$\sigma$ upper limit is S$_{1.4} <
0.15~\mu$Jy. The optical spectrum published by 
(Peroux et al. 2001) shows this
to be a normal emission line quasar.

{\bf PSS J1347+4956} This source is clearly detected with S$_{1.4} =
91\pm 20 \mu$Jy at (J2000) $\rm 13^h 47^m 43.30^s$ $+49^o 56' 21.6''$,
within 0.3$''$ of the optical QSO position. 
A Gaussian fit results in a marginally resolved source with a
deconvolved FWHM $\sim 1''$, with a comparable uncertainty. The source
is not detected on a 5 GHz image with an rms noise of 32$\mu$Jy at
0.5$''$ resolution. The optical spectrum of this source
is not yet published.

{\bf PSS J1418+4449} A radio continuum source 
with S$_{1.4} = 51\pm16~\mu$Jy is detected at 
(J2000) $\rm 14^h 18^m 31.69^s$ $+44^o 49' 37.5''$,
within 0.2$''$ of the optical  QSO position.
The optical spectrum of this source is not yet published. 

{\bf PSS J1554+1835}  This source was not detected at 1.4 GHz,
with a 3$\sigma$ upper limit of S$_{1.4} < 51~\mu$Jy. 
The optical spectrum of this source is not yet published.

{\bf PSS J1646+5514} This source is not detected  at 1.4 GHz,
with a 3$\sigma$ upper limit of S$_{1.4} < 90~\mu$Jy. 
The optical spectrum of this source is not yet published.

{\bf PSS J2322+1944} This source has the highest S$_{250}$ 
in the recent survey of Omont et al. (2001). 
It is clearly detected at 1.4 GHz
with at total flux density of $98\pm20 \mu$Jy, 
and a peak surface brightness of $66\pm15 \mu$Jy beam$^{-1}$ at
(J2000) $\rm 23^h 22^m 07.2^s$ $+19^o 44' 23.07''$, within 
0.1$''$ of the optical QSO position.  The source
is spatially resolved, with a nominal FWHM = 1.5$''$
derived from Gaussian fitting. At 5 GHz the source is not detected
at full resolution with an rms noise of 30$\mu$Jy.
Convolving to the resolution of the  1.4 GHz image, we can set a 
2$\sigma$ lower limit to the spectral index between 1.4 and 5 GHz of  
--0.1. The optical spectrum of this source is not yet published.

\section{Analysis}

An important question which these radio observations address is
whether the 250 GHz emission is synchrotron radiation, ie. a radio
loud AGN component, or thermal dust emission.  For most of the sources 
in the survey S$_{250}$ is  larger than S$_{1.4}$
by at least two orders of magnitude, implying
a sharply rising spectrum between 1.4 and 250 GHz of index +1 or
greater. This would be highly unusual for the integrated spectrum of a
radio loud AGN, even in the case of a synchrotron self absorbed
source, for which flat spectra into the mm are occasionally seen
(Sanders et al. 1989; Owen, Spangler, and Cotton 1980).  
Even in the cases of the two brightest radio sources in the
sample, J1048+4407 and J1057+4555,  S$_{250}$ is a factor five to ten
larger than S$_{1.4}$.  Moreover, observations at 5 GHz of the
four brightest sources at 1.4 GHz in this sample, 
including J1048+4407 and J1057+4555, indicate falling
spectra at cm wavelengths, with no detection in any case of a
compact, flat or rising spectrum 
AGN radio component.  Overall, these data argue strongly that
the 250 GHz emission is thermal emission 
from warm dust, and not synchrotron
radiation powered by the AGN.  A similar conclusion was reached for
250 GHz detected QSOs from the Sloan Digital Sky Survey (SDSS), based
on 1.4 GHz and 43 GHz observations (Carilli et al. 2001a), and by Yun et
al. (2000) for the  BR QSOs in the sample of Omont et al. (1996a).
Further strong supporting evidence that the 250 GHz emission from
these sources is thermal emission from warm dust comes from the 
fact that many of these sources have steeply rising spectra between
250 GHz and 350 GHz (Isaak et al. 2001; see Table 1). 

Miller et al. (1990) suggested a division between radio loud and
radio quiet QSOs at a rest frame 5 GHz spectral luminosity
of  $10^{26}$ W Hz$^{-1}$. This corresponds to
a flux density of S$_{1.4} \sim 1$ mJy for a source
at $z = 4.2$ assuming a spectral index of --0.8.
According to this criterion, J1057+4555 and J1048+4407 can
be considered radio-loud, and the rest of
the sources are clearly radio quiet. 
The detection of two radio loud sources in our
sample of 12 is consistent with the 10$\%$ fraction of
radio loud QSOs found in optically selected samples of QSOs as a
whole (White et al. 2000, Stern et al. 2000, Kellermann et al. 1989;
Hooper et al. 1995). 

Stocke et al. (1992) consider the radio properties 
of QSOs as normalized by their optical  luminosities. They 
define the R parameter as the ratio of the rest frame
5 GHz spectral luminosity to the blue spectral luminosity.
For high blue luminosity QSOs 
they define a radio loud source as  $\rm R > 10$, and
a radio quiet source as $\rm R < 1$. Table 1 shows the R parameter
in Column 6. Stocke et al. (1992) suggest a bi-modal radio luminosity
function, with a significant deficit of sources in the range $\rm 10
\ge R \ge 1$. However, the reality of this deficit has been
called into question recently by White et al. (2000; cf. Stern
et al. 2000), and Stocke et al. (1992) show that the deficit is less
clearly delineated
for high  luminosity QSOs relative to lower luminosity sources. 
In general, these definitions apply to radio emission that is
powered by the AGN itself. In this paper we are most interested
in radio emission at a much lower level, corresponding to emission
associated with a possible starburst co-eval with the AGN. 

\subsection{Comparison with Star Forming Galaxies}

A more relevant comparison for the purposes of our study is
with the expected radio luminosities of star forming galaxies.
Figure 3 shows the relationship between redshift and 
the 250-to-1.4 GHz spectral
index for a star forming galaxy taken from 
the model of Carilli \& Yun (2000) based on 17 low redshift galaxies
(roughly equivalent to a modified black body spectrum 
with $\rm T_D = 50 K$ and dust emissivity index = 1.5). 
The relationship relies on the tight radio-to-far IR correlation seen
for star forming galaxies in the nearby universe (Condon 1992).
The dotted curve gives the rms scatter for the 17 galaxies.
The symbols are the results for the  QSOs in Table 1. 
The arrows  indicate non-detections in the
radio, and hence lower limits to the spectral index.  On this diagram, 
a point located below the curve would indicate  a
source that is radio-loud relative to the 
standard radio-to-far IR correlation for star forming galaxies, 
while  a point located above the curve 
would indicate a source which is radio-quiet relative to this
relationship. 

The two radio loud AGN  (J1048+4407 and J1057+4455) are clearly evident as
being well below the star forming galaxy curve. 
The fainter radio detections are all within the 
1$\sigma$ scatter defined by star forming galaxies, while the upper
limits are consistent with the range defined by star forming 
galaxies. 

\subsection{Spectral Energy Distributions}

Figure 4 shows the mean radio-through-optical SEDs for radio-loud and
radio-quiet QSOs derived from observations of the Palomar-Green (PG)
sample of QSOs determined by Sanders et al. (1989).  The solid curve
shows the SED for radio loud QSOs, while the dashed curve shows the
SED for radio quiet QSOs. The hatched regions indicate the scatter in
the measured values for the PG sample.  Note that the QSOs in the PG
sample are typically at lower redshift ($z \le 2$), and are an order
of magnitude less luminous in the rest-frame UV than the high redshift
sources in Table 1. Most of the PG sources are 
in the range of $\rm M_B = -23$ to $-27$.

The data points in Figure 4 show the results for the sources 
in Table 1,  normalized  by their blue spectral
luminosities. The normalized mm and submm data for all
the sources  fall within the range defined by the PG sample.
Considering the 1.4 GHz data in Figure 4, the two radio loud QSOs 
(J1048+4407 and J1057+4555) are clearly evident 
in the normalized SEDs in Figure 4, situated about  an order of
magnitude  above the radio quiet regime. 
The weaker radio detections, and radio upper limits,
lie within a factor of three of the radio quiet SED.

\section{Discussion}

A fundamental implication of these cm wavelength observations
of mm-emitting  high redshift QSOs is that the 
mm wavelength emission in all cases is almost certainly thermal
dust emission, and not synchrotron emission 
from a flat spectrum radio loud AGN. In no case do we detect
a compact, flat spectrum AGN radio component that could
explain the mm continuum emission. 

The importance of pushing to limits of 10's of $\mu$Jy at 1.4 GHz for
high redshift  sources which show thermal emission from 
warm dust is that we probe well below the
radio-loud AGN luminosity regime, into the regime expected for active
star forming galaxies.  We detect 7 sources at 1.4 GHz.
For 5 of these sources 
the ratios of 1.4 GHz to 250 GHz spectral luminosities
are consistent with those  expected for
active star forming galaxies  based on the tight radio-to-FIR
correlation for such systems.  
For  the remaining two sources the
radio continuum luminosities and morphologies indicate steep spectrum,
radio loud emission from a jet-driven radio source.
The 5 radio non-detections are all consistent with star forming
galaxies, although we cannot rule out the possibility that the sources
are radio quiet relative to a typical star forming galaxy.

A flux density of 50 $\mu$Jy at an observing frequency of
1.4 GHz implies a rest frame spectral luminosity of $5.3\times
10^{24}$ W Hz$^{-1}$ at 1.4 GHz at $z = 4.2$, assuming a spectral index
of --0.8. For comparison, the $z = 0.019$ nuclear starburst galaxy 
Arp 220 has a rest frame spectral luminosity of $4.7\times 10^{23}$ W
Hz$^{-1}$, while that for the $z = 0.0044$ low luminosity
(Fanaroff-Riley Class I; FR I) radio jet source M87 is $1.8\times
10^{25}$ W Hz$^{-1}$, and that for the $z = 0.057$, luminous (FR II)
radio galaxy Cygnus A is $2.1\times 10^{28}$ W Hz$^{-1}$.

Arcsecond resolution radio observations provide constraints on source
sizes unavailable from single dish mm observations.  An important
point is that, unless the dust temperatures are much larger than 50 K,
the emitting regions have absolute lower limits to their sizes of
$\sim 0.1''$, set by assuming optically thick dust emission (Carilli
et al. 2001a). This is an extreme lower limit, given that the emission
is most likely optically thin. The two sources J0808+5215 and J1418+4449
are too faint for meaningful limits to be placed on source sizes.  The
brighter sources J1347+4956 and J2322+1944 both appear resolved, with
sizes $\sim 1''$.  If the dust emission has a similar spatial
distribution as the radio continuum emission then this argues against
a single, small ($\sim 0.1'' = 0.6$kpc), high surface brightness
region as the origin for the dust emission, as would be expected if
the dust was heated locally by the AGN.  Such spatially 
extended emission on kpc-scales is 
interesting when seen in comparison with luminous star forming
galaxies in the low $z$ universe, ie. the Ultra-Luminous Infrared
Galaxies (ULIRGs; $\rm L_{IR} \ge 10^{12}$ L$_\odot$). The infrared
and radio continuum emission from low $z$ ULIRGs typically show most
of the emission coming from a compact, nuclear starburst region, with
a scale of only a few hundred pc (Downes and Solomon 1999, Condon
1992). Hence, the larger spatial extent of the emitting regions
in  these two sources suggests a fundamental difference
between these sources and low-$z$ ULIRGs.
A similar conclusion has been reached for a few
other high redshift dust and CO emitting sources (Papadopoulos et
al. 2001, Papadopoulos et al. 2000, Richards 2000, Kohno et al. 2001).

The blue-normalized optical-through-radio SEDs for these sources are
within the broad range delineated by the lower redshift
PG sample (Figure 4). The typical UV luminosity
(which dominates the bolometric luminosity)
for the QSOs in Table 1 is $\sim 10^{14}$ L$_\odot$, 
while the typical IR luminosity is about $\sim 10^{13}$ L$_\odot$,
assuming a standard dust spectrum (Omont et al. 2001). Hence, the IR
emission constitutes $\sim 10\%$ of the bolometric luminosity for
these galaxies, similar to low redshift, lower luminosity QSOs. 
This implies that  we are not seeing a
dominant new broad-band spectral emission `feature' at
mm wavelengths in high redshift 
QSOs relative to the normalized SEDs of lower redshift QSOs. 

The fact that the radio-to-FIR SEDs of most of the sources in Table 1
are consistent with those expected for star forming galaxies
is by no means proof that the sources are actively
forming stars. The important point is that the cm-to-submm 
SEDs of radio quiet PG QSOs are
also consistent with the radio-to-FIR correlation for
star forming galaxies (Sanders et al. 1989).
Sanders et al. (1989) consider whether this implies
star formation in low redshift QSOs as well, and conclude
that the agreement is merely coincidental. A similar
statement could be made for the high redshift sources.

Omont et al. (2001) consider in detail the implications of 
the detection of thermal emission from dust in high redshift QSOs in
the context of co-eval AGN and starbursts in high redshift QSOs, as
suggested by the recently discovered correlation between black hole
mass and stellar bulge mass in low redshift galaxies. We re-address
these issues briefly herein (see also Isaak et al. 2001; Priddey \&
McMahon 2001).  We adopt the relation for 
low redshift galaxies of~ $\rm {{M_{bulge}}\over{M_{BH}}} = 160$,
as derived by Magorrian et al. (1998).
If we assume that  black hole and 
stellar bulge formation are co-eval, then:
$$\rm  {{\dot{M}_{bulge}}\over{\dot{M}_{BH}}} \sim
{{M_{bulge}}\over{M_{BH}}} = 160 ~~~~~~(1) $$ 
The  black hole luminosity due to accretion is
given by (Begelman, Blandford, and Rees 1986):
$$\rm L_{acc} = 1.3\times 10^{13} \epsilon {\dot{M}_{BH}}~ L_\odot ~~~~~~(2)$$
where $\epsilon$ is the mass-to-energy conversion efficiency, and $\rm
{\dot{M}}$ is in M$_\odot$ year$^{-1}$.
The relationship between IR luminosity and 
star formation rate has been derived by numerous authors
(see Kennicutt 1998). We use
the most recent derivation given in Carilli et al. (2001b):
$$\rm L_{IR} = 4\times 10^9  {\dot{M}_{bulge}}~ L_\odot~~~~~~~~(3)$$
For a  star formation driven IR luminosity of 
10$^{13}$ L$_\odot$, typical for the sources discussed herein,
equation 3 implies $\rm {\dot{M}_{bulge}} = 2500$ M$_\odot$
year$^{-1}$. Using this value in equation 1 then leads to
$\rm {\dot{M}_{BH}} = 16$ M$_\odot$ year$^{-1}$.
Using this value of $\rm {\dot{M}_{BH}}$ in equation 2, and using 
a typical luminosity due to accretion onto the black hole
for this sample of 10$^{14}$ L$_\odot$ 
(ie. the QSO optical-through-UV luminosity) 
yields  $\rm \epsilon = 0.5$.
A final constraint comes by assuming the QSO luminosity
is Eddington limited: 
$\rm L_{acc} \le L_{Edd} = 3.3\times 10^{12} {M_8}$ L$_\odot$, 
where $\rm M_8 = {{M_{BH}}\over{10^8 M_\odot}}$.
For  $\rm L_{acc} = 10^{14}$ L$_\odot$ this
implies $\rm M_{BH} \ge 3\times 10^9$ M$_\odot$. 
The above relations are consistent with the formation
of a  $3\times 10^9$ M$_\odot$ black hole, and a
$5\times 10^{11}$ M$_\odot$ spheroidal galaxy, in
$2 \times 10^8$ years. The main problem with this
simple calculation is the high
efficiency (50$\%$) required for conversion of accreted mass into
black hole luminosity, as pointed out
and discussed at length by Omont et al. (2001). 

The issue of co-eval black hole and stellar bulge formation at high
redshift remains open. The data presented herein are consistent with
such an idea, but certainly do not constitute proof thereof.
There are a number of critical observations that can be made to
address this interesting question. First is sub-arcsecond resolution
imaging at cm and mm wavelengths of the sources to determine the
spatial distribution of the emitting regions. Second is a search for
large reservoirs of molecular gas through CO observations. And third
is sub-arcsecond resolution imaging at optical and near IR wavelengths
to study the stars in the parent galaxies. All of these are at the limit
of what can be done with current instrumentation, but will become
routine with future instruments such as the Atacama Large Millimeter
Array, the Expanded VLA, and the Next Generation Space Telescope. 

\vskip 0.2truein 

The VLA is a facility of the National Radio
Astronomy Observatory (NRAO), which is operated by Associated
Universities, Inc. under a cooperative agreement with the National
Science Foundation.
This work was based on observations carried out with the IRAM 30m
telescope.  IRAM is supported by INSU/CNRS (France), MPG (Germany) and
IGN (Spain).  CC acknowledges support from the Alexander
von Humboldt Society. 


\newpage
 
\begin{table}
\caption{High Redshift QSOs} 
\vskip 0.2in
\begin{tabular}{ccccccc}
\hline
\hline
Source & $z$ & S$_{250~ \rm GHz}$ &
S$_{1.4~ \rm GHz}$ & S$_{350 ~ \rm GHz}$ & M$_{\rm B}$ & R$^a$ \\
Source & ~  & mJy & mJy & mJy & ~ & ~ \\
\hline
PSS J0808+5215 & 4.45 & $6.6\pm0.6$ & 0.058$\pm$0.018d &
17.4$\pm$2.8 & --28.7  & 0.29 \\
PSS J1048+4407 & 4.40 & $4.6\pm0.4$ & 0.430$\pm$0.020d &
12.0$\pm$2.2 & --27.4 & 6.90 \\
PSS J1057+4555 &  4.12 &  $4.9\pm0.7$  & 1.24$\pm$0.04d &
19.2$\pm$2.8 & --29.4 &  2.75 \\
APM B1117--1329 & 3.96 & $4.1\pm0.7$  & 0.022$\pm$0.030 &
27$\pm$13 & --28.2 &  $<$0.33 \\ 
APM B1144--0723 & 4.15 &  $6.0\pm0.9$ & 0.046$\pm$0.040 & 
7$\pm$2 & --27.7 & $<$0.39 \\ 
PSS J1248+3110 & 4.35 & $6.3\pm0.8$ & 0.071$\pm$0.015d &
12.7$\pm$3.4 & --27.6  & $<$0.39 \\
PSS J1253--0228 & 4.00 & $5.5\pm0.8$ & --0.043$\pm$0.050 & ~ &
--27.2  & $<$1.58 \\ 
PSS J1347+4956 & 4.56 &  $5.7\pm0.7$ & 0.091$\pm$0.020d &
8.5$\pm$3.9 & --28.6 &  0.50 \\
PSS J1418+4449 & 4.32 & $6.3\pm0.7$ & 0.051$\pm$0.016d &
10.4$\pm$2.3 & --28.6 & 0.26 \\ 
PSS J1554+1835 & 3.99 & $6.7\pm1.1$ & 0.018$\pm$0.017 & ~ &
--26.6 &  $<$0.93 \\
PSS J1646+5514 & 4.10 & $4.6\pm1.5$ & 0.022$\pm$0.030 &
9.5$\pm$2.5 & --28.7 & $<$0.45 \\
PSS J2322+1944 & 4.11 &  $9.6\pm0.5$ & 0.098$\pm$0.015d &
22.5$\pm$2.5 & --28.2 &  0.18 \\
\hline
\end{tabular}

~$^a$See section 4 for the definition of R. Upper limits are 
3$\sigma$ for radio non-detections. 

\end{table}

\clearpage
\newpage

\centerline{Figure Captions}

\noindent {\bf Figure 1}: 
Images of the 12 QSOs from Table 1 
at 1.4 GHz. The FWHM of the Gaussian restoring 
CLEAN beam is shown as an inset to each figure.
Contour levels are a geometric progression is
$2^{1\over2}$. The first level is 2$\sigma$
in each case, with $\sigma$ = rms noise on each image, 
as specified below. The cross on each image indicates the 
position of the optical QSO as given in Table 1. 
{\bf Page 1:} Upper left: J0808+5215, $\sigma = 18 \mu$Jy. 
Upper right: J1048+4407,  $\sigma = 20\mu$Jy.
Lower left: J1057+4555,   $\sigma = 15\mu$Jy.
Lower right: B1117--1329,   $\sigma = 30\mu$Jy.  
{\bf Page 2:} Upper left: B1144--0723, $\sigma = 40 \mu$Jy.
Upper right: J1248+3110,  $\sigma = 15\mu$Jy.
Lower left: J1253--0228,   $\sigma = 50\mu$Jy.
Lower right: J1347+4956,   $\sigma = 20\mu$Jy.  
{\bf Page 3:} Upper left: J1418+4449, $\sigma = 16 \mu$Jy.
Upper right: J1554+1835,  $\sigma = 17\mu$Jy.
Lower left: J1646+5514, $\sigma = 30\mu$Jy.
Lower right: J2322+1944,   $\sigma = 15\mu$Jy.

\noindent {\bf Figure 2}:
Images of J1048+4407 (upper) and J1057+4555 (lower) 
at 5 GHz.  The FWHM of the Gaussian restoring 
beam is  $0.86'' \times 0.41''$
with a major axis position angle =  --70$^o$
for  J1048+4407, while that for J1057+4555
is $0.77'' \times 0.40''$
with a major axis position angle =  --74$^o$.
Contour levels are a geometric progression is
$2^{1\over2}$. The first level is 2$\sigma$,
with $\sigma = 26 \mu$Jy. 
The cross on the  J1048+4407 image 
shows the position of the  optical QSO. 
The crosses on the J1057+4555 image show the positions
of the peak surface brightness for the northeast 
and southwest radio lobes, and the optical QSO (center).

\noindent {\bf Figure 3}:
The solid curve shows the relationship between 
redshift and the observed spectral index between 1.4 and 250 GHz
for star forming galaxies, as derived from the models presented in 
Carilli \& Yun (2000). The dotted lines show the rms scatter in the
distribution. The solid symbols show data for high redshift QSOs
from Table 1.  The squares are for sources
detected at 1.4 GHz, while the arrows show lower limits (2$\sigma$)
to the spectral indices for sources that were not detected at 1.4
GHz. 

\noindent {\bf Figure 4}:
The curves show the
radio through UV Spectral Energy Distributions (SEDs) for the
Palomar-Green QSOs, taken from Sanders et al. 1989.
All data have been normalized to the rest frame blue spectral luminosity. 
For frequencies above
$3\times10^{12}$ Hz, the solid curve shows the mean spectral energy
distribution for the PG QSO sample, which is 
approximately the same
for radio-loud and radio-quiet sources. Below
$3\times10^{12}$ Hz, the hatched regions
show the typical ranges given the scatter in the
observed properties of the PG QSO sample at cm and mm wavelengths,
with the solid line delineating the radio-loud sources and
the dashed-line the radio quiet sources. 
The solid symbols show the values derived for the QSOs 
listed in Table 1.
The arrows show (2$\sigma$) upper limits at 1.4 GHz. 

\clearpage
\newpage

\begin{figure}
\hskip -1.0in
\psfig{figure=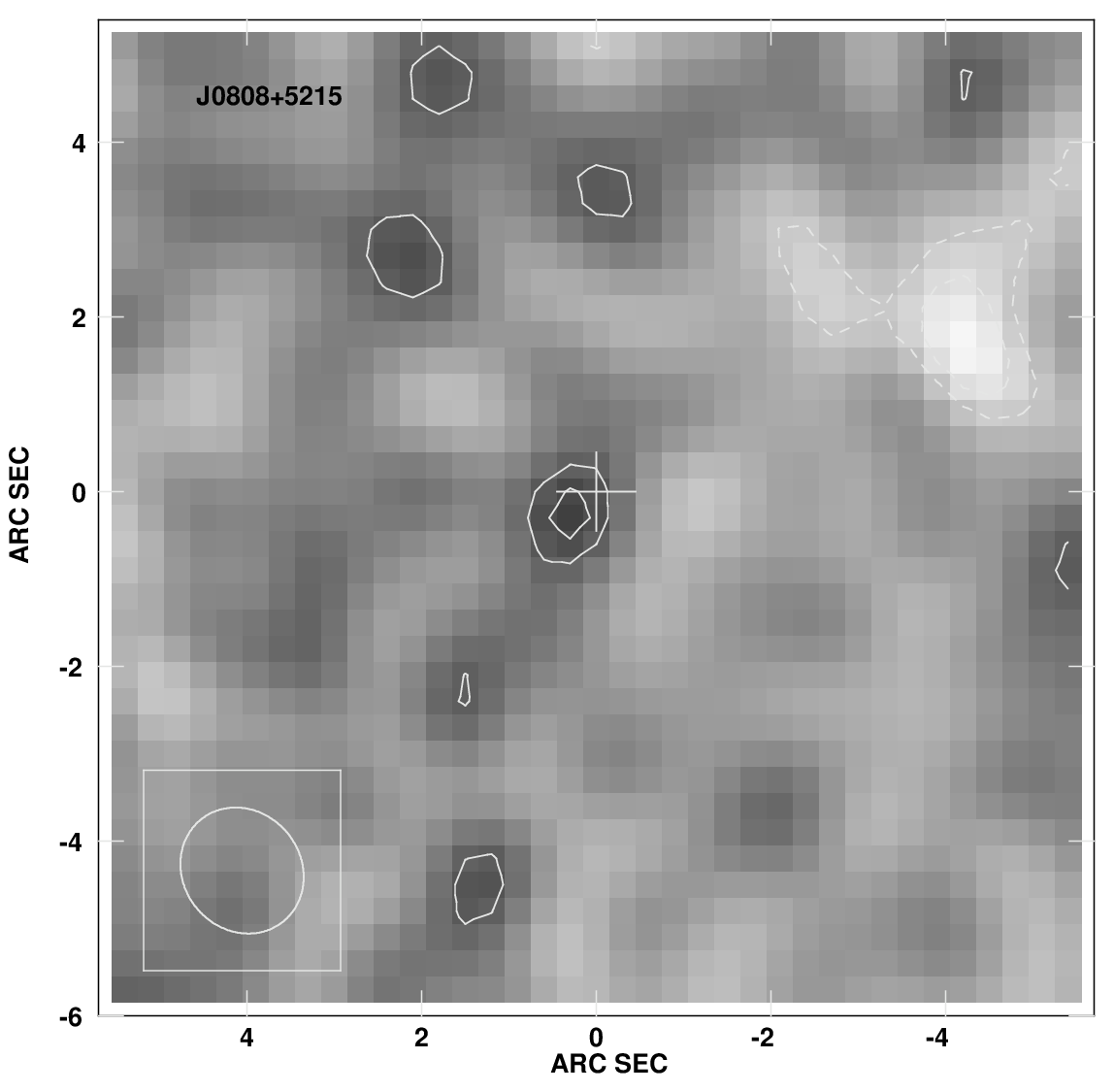,width=4in,angle=-90}
\vskip -3.95in
\hspace*{3.0in}
\psfig{figure=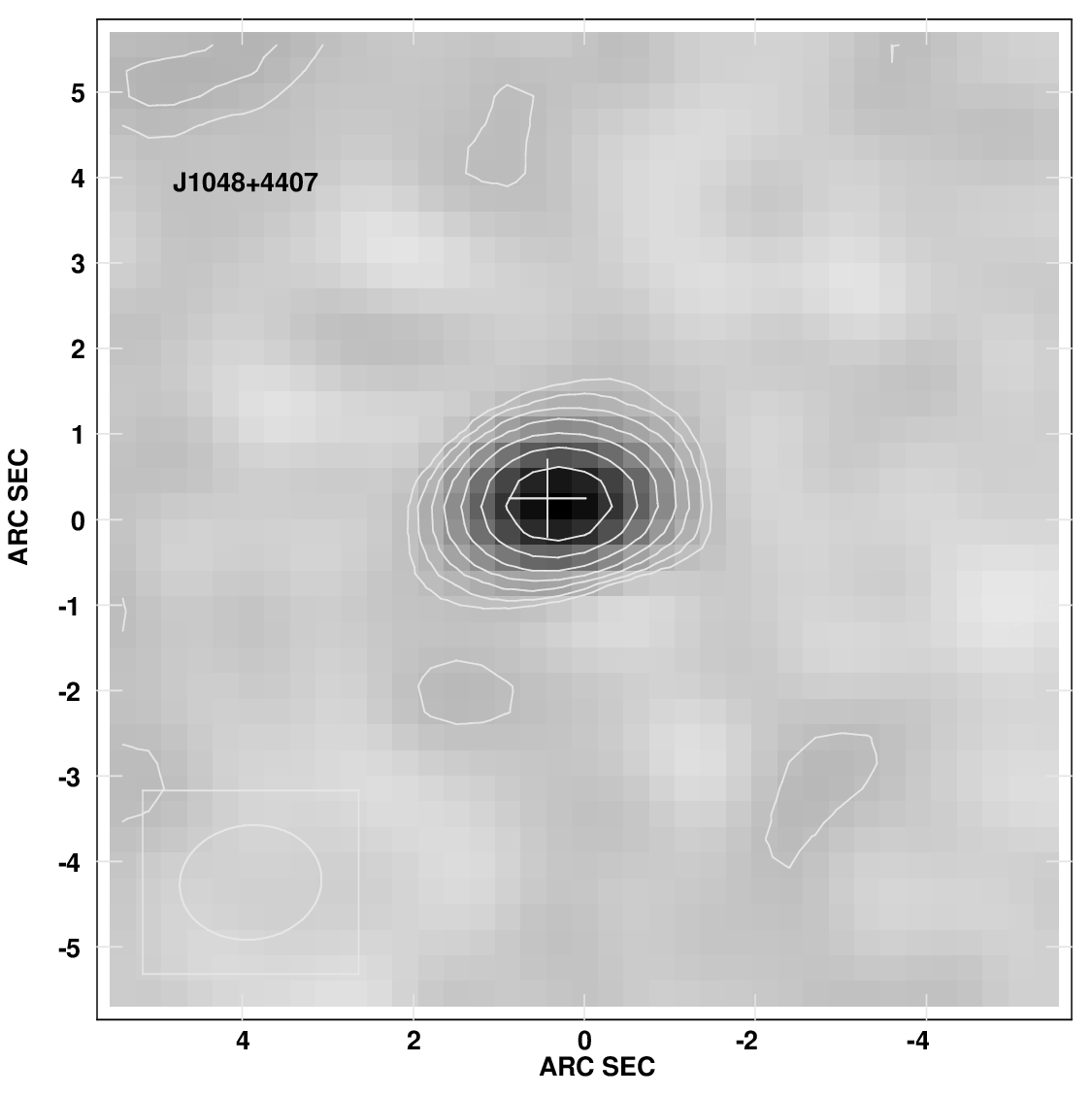,width=4in}
\hspace*{-1.05in}
\psfig{figure=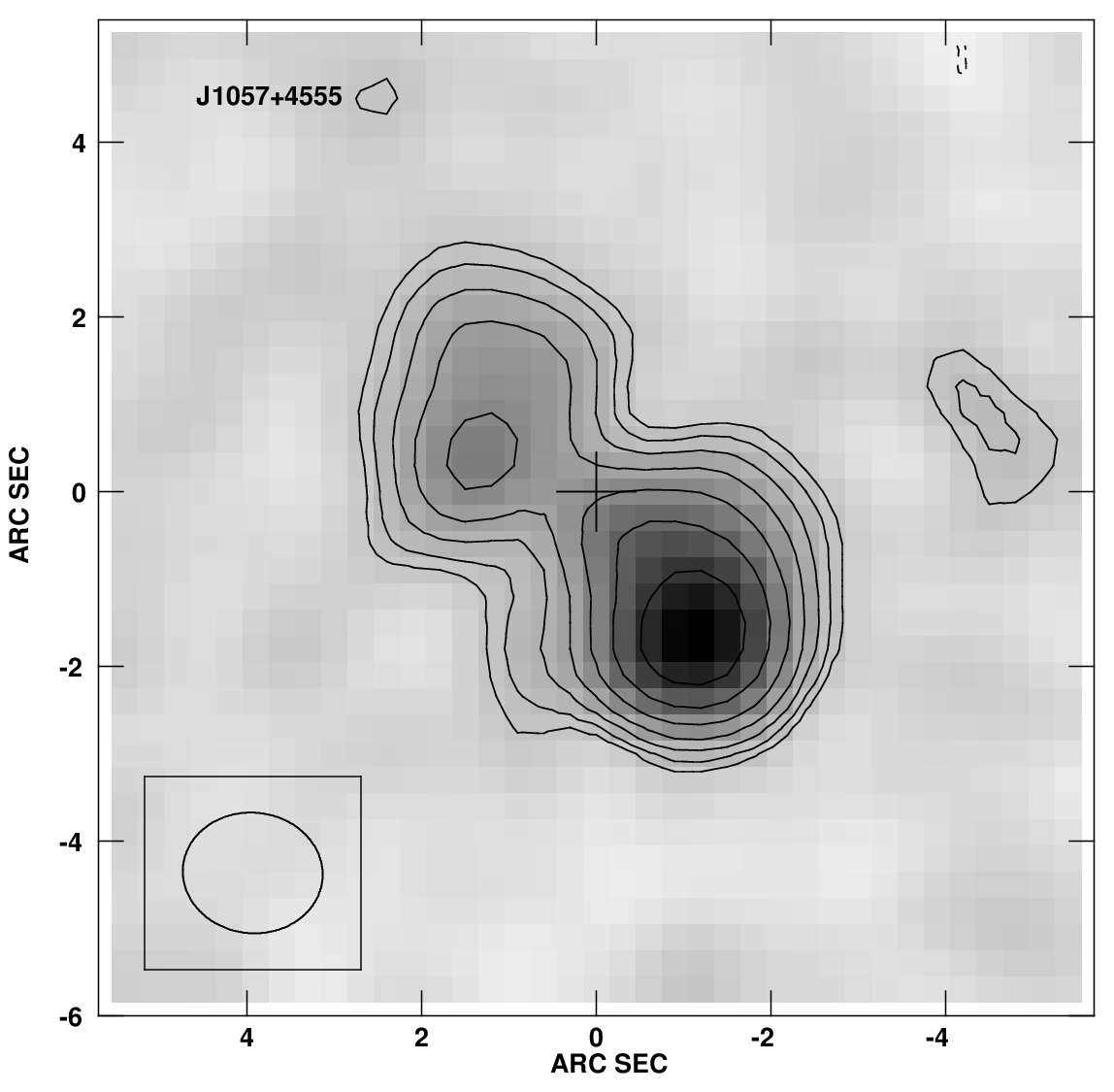,width=4in,angle=-90}
\vskip -4.2in
\hspace*{3.0in}
\psfig{figure=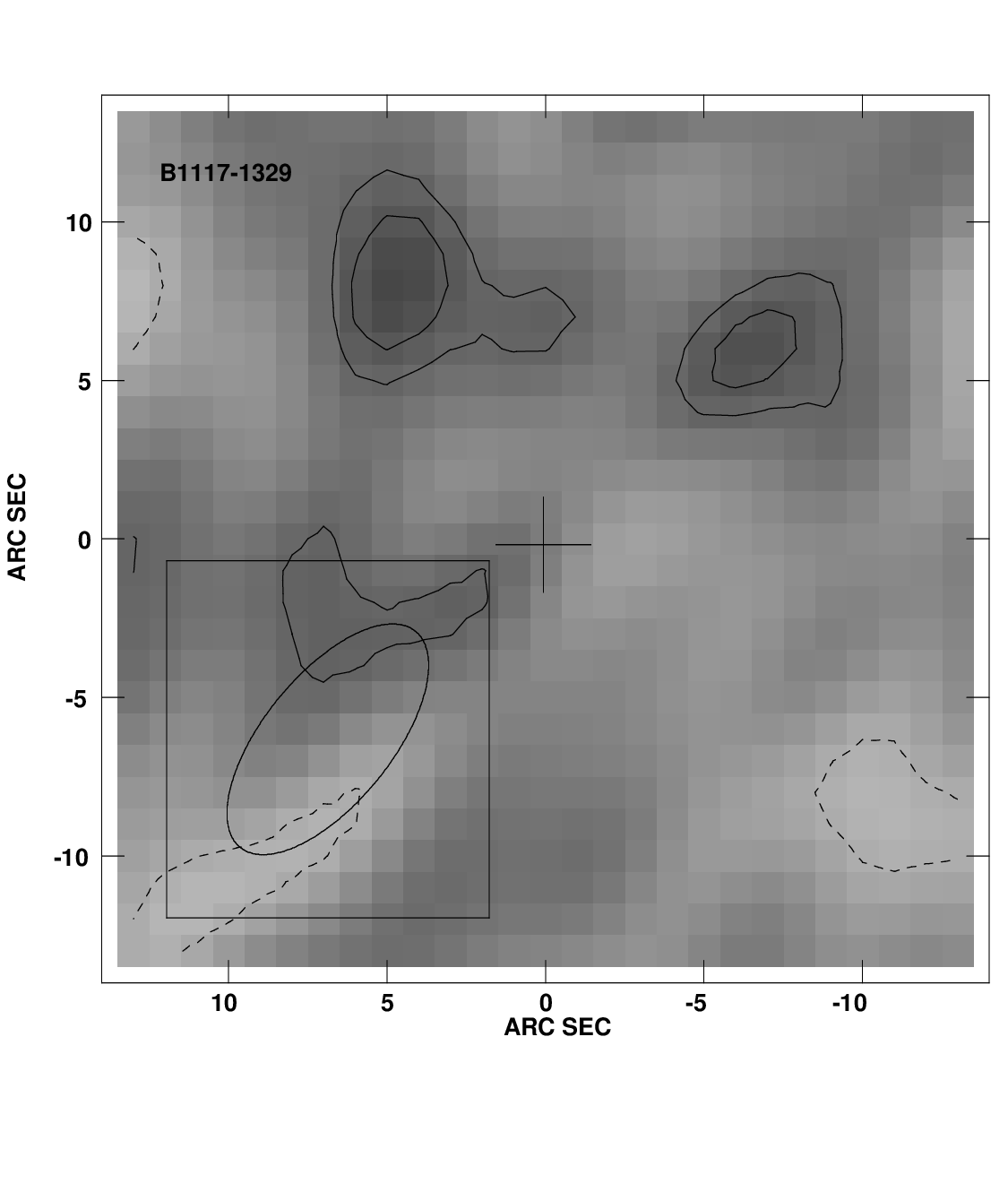,width=4in}
\caption{}
\end{figure}

\clearpage
\newpage

\begin{figure}
\hskip -1.0in
\psfig{figure=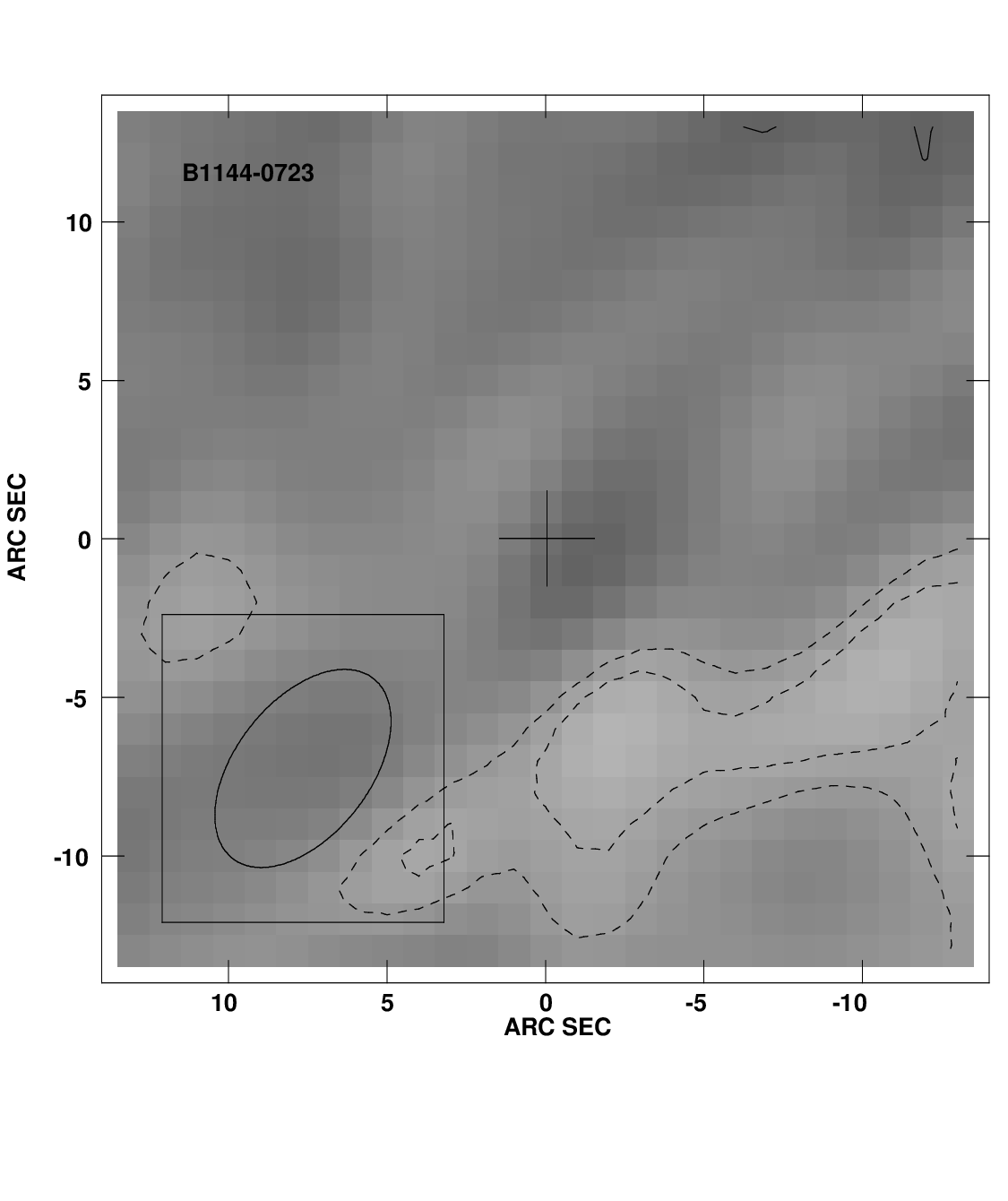,width=4in}
\vskip -4.5in
\hspace*{3.0in}
\psfig{figure=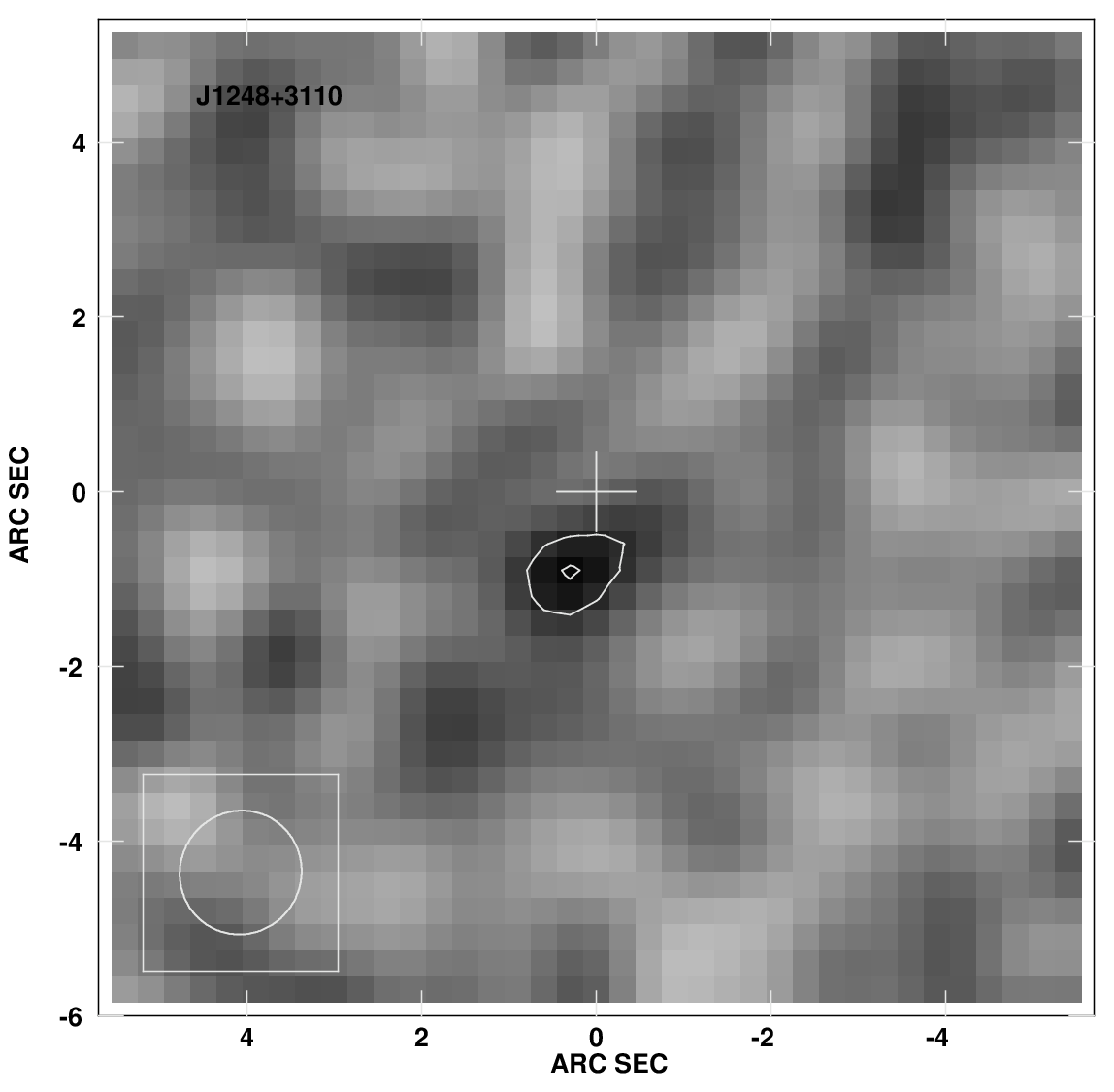,width=4in,angle=-90}
\hspace*{-1.05in}
\psfig{figure=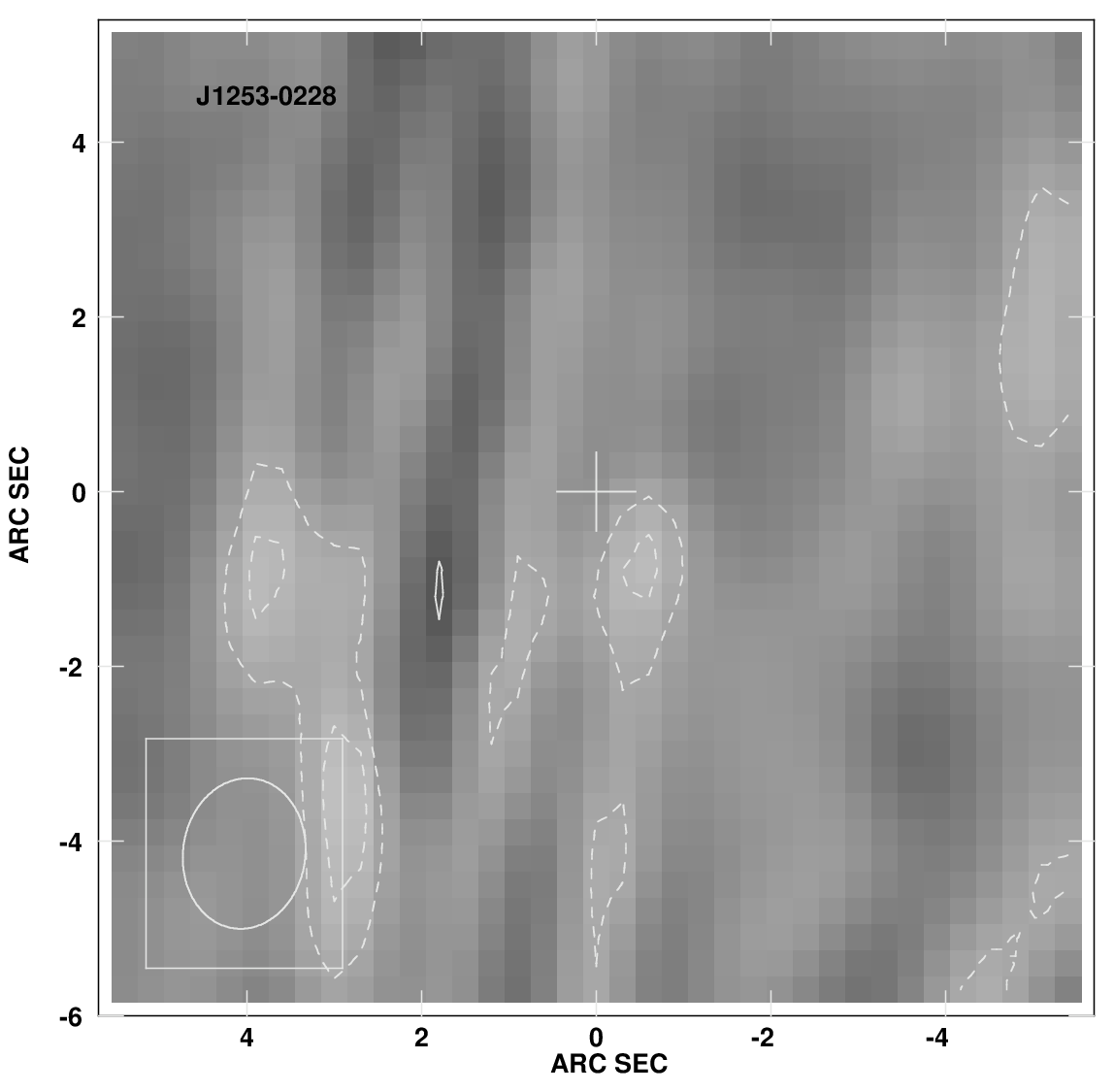,width=4in,angle=-90}
\vskip -3.95in
\hspace*{3.0in}
\psfig{figure=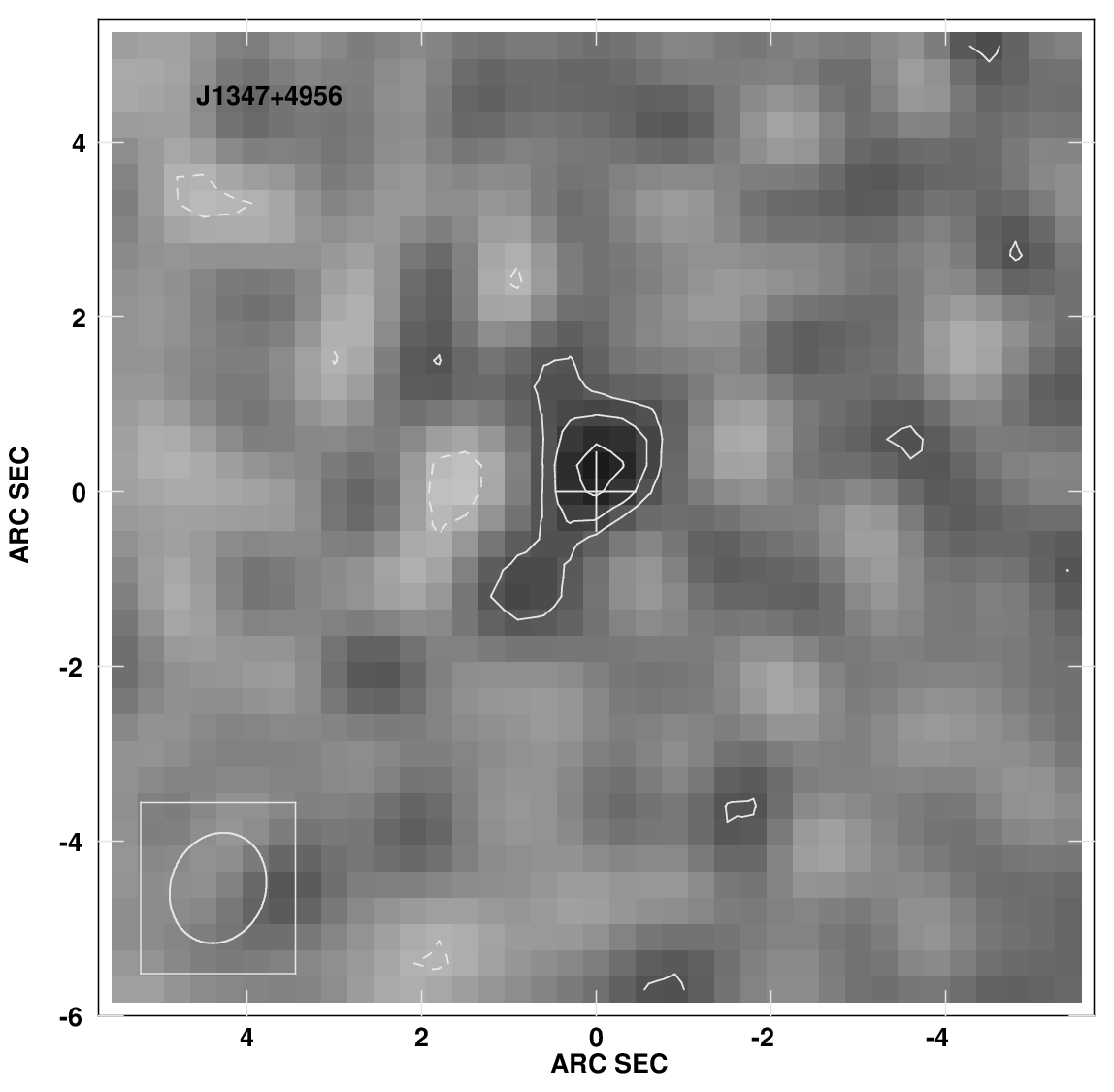,width=4in,angle=-90}
\end{figure}
 
\clearpage
\newpage

\begin{figure}
\hskip -1.0in
\psfig{figure=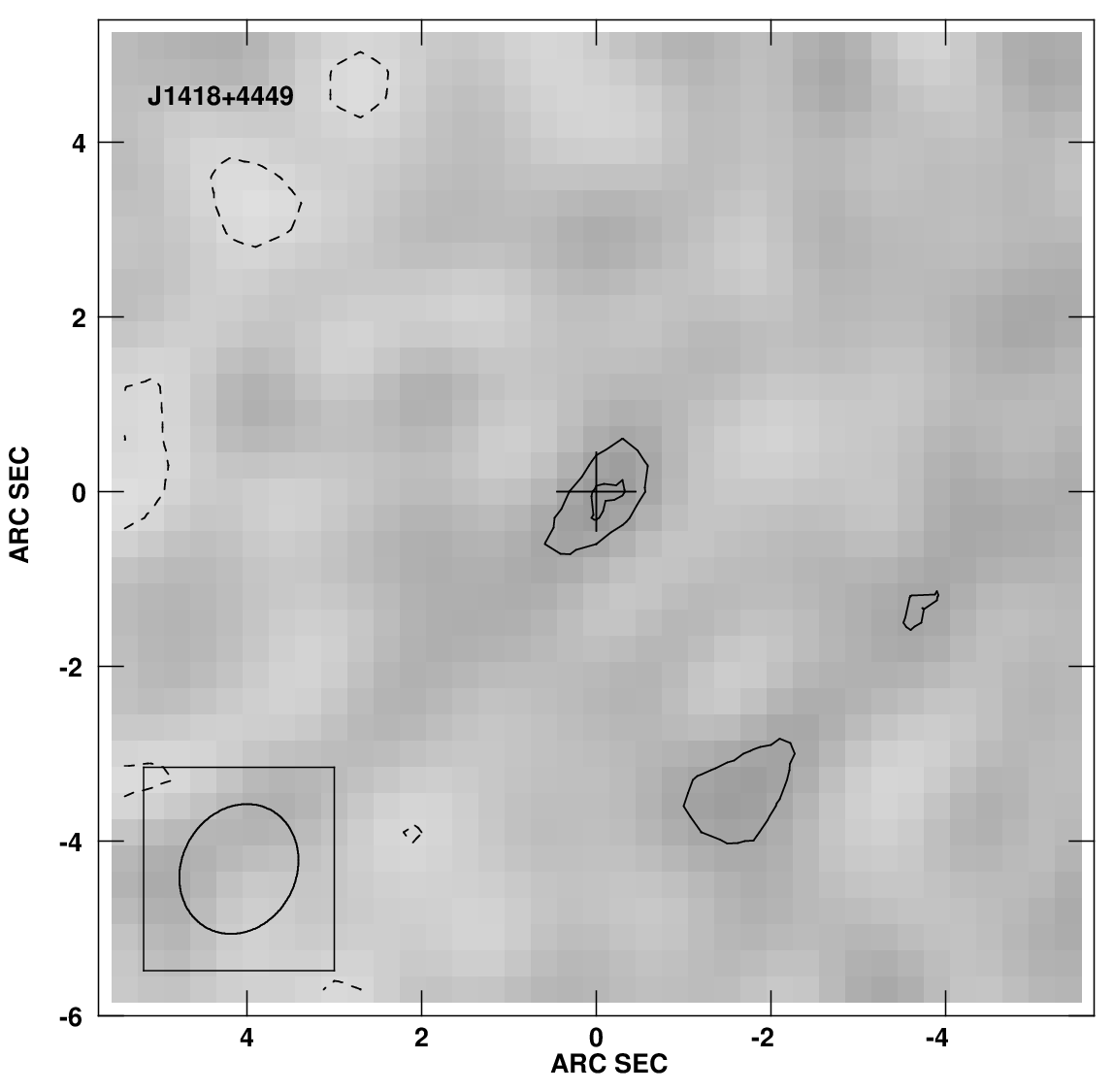,width=4in,angle=-90}
\vskip -3.95in
\hspace*{3.0in}
\psfig{figure=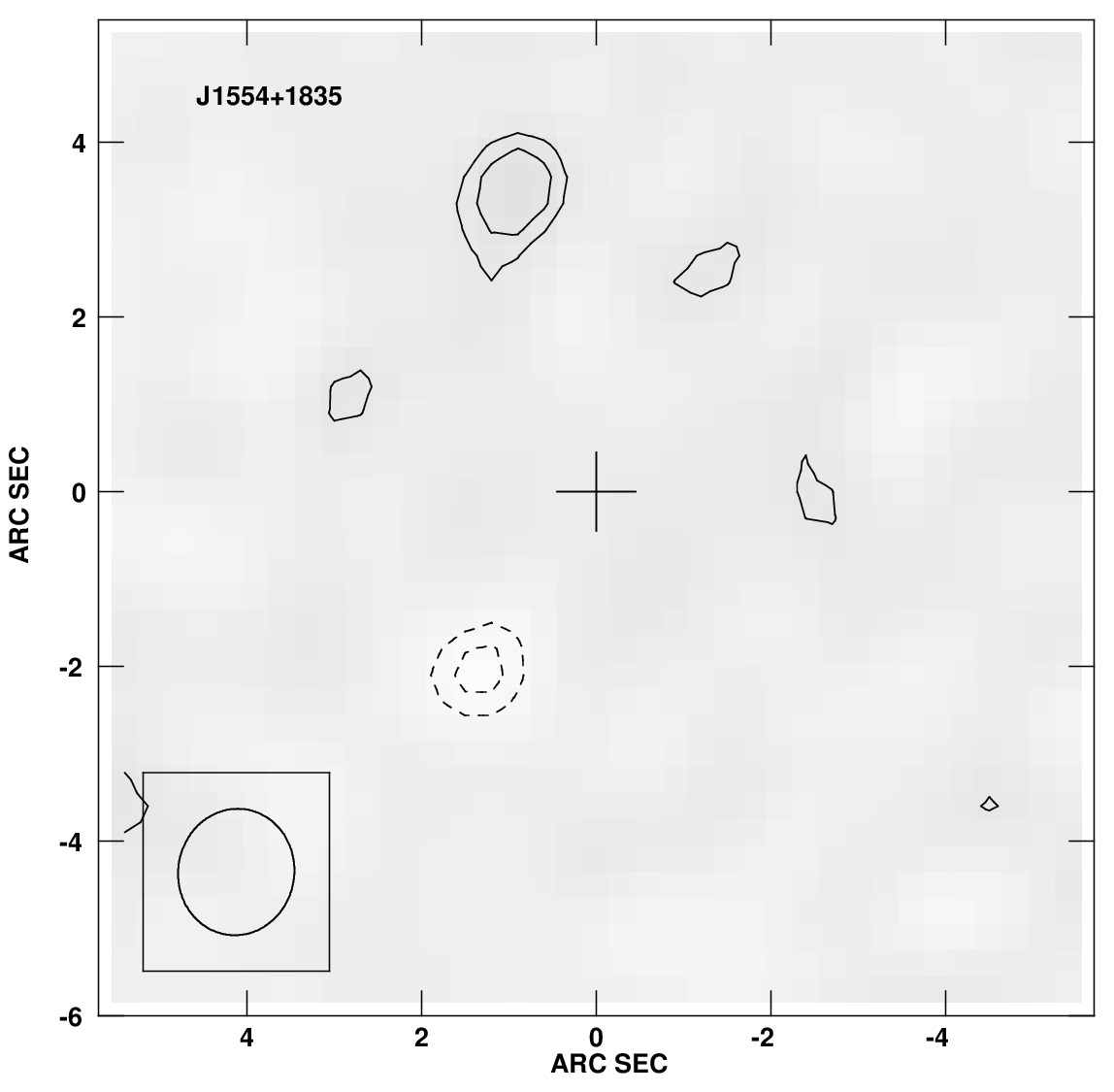,width=4in,angle=-90}
\hspace*{-1.05in}
\psfig{figure=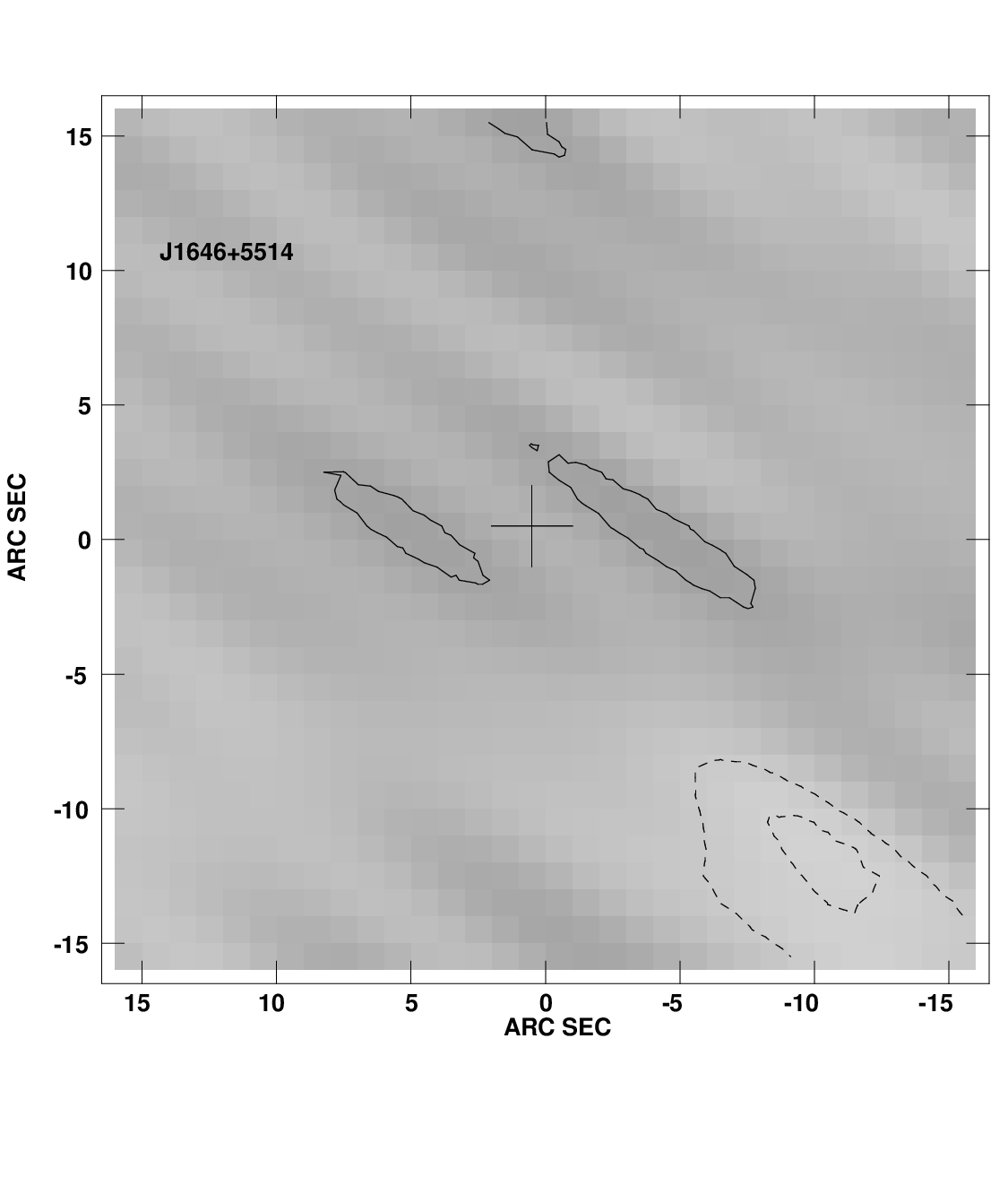,width=4in}
\vskip -4.35in
\hspace*{3.0in}
\psfig{figure=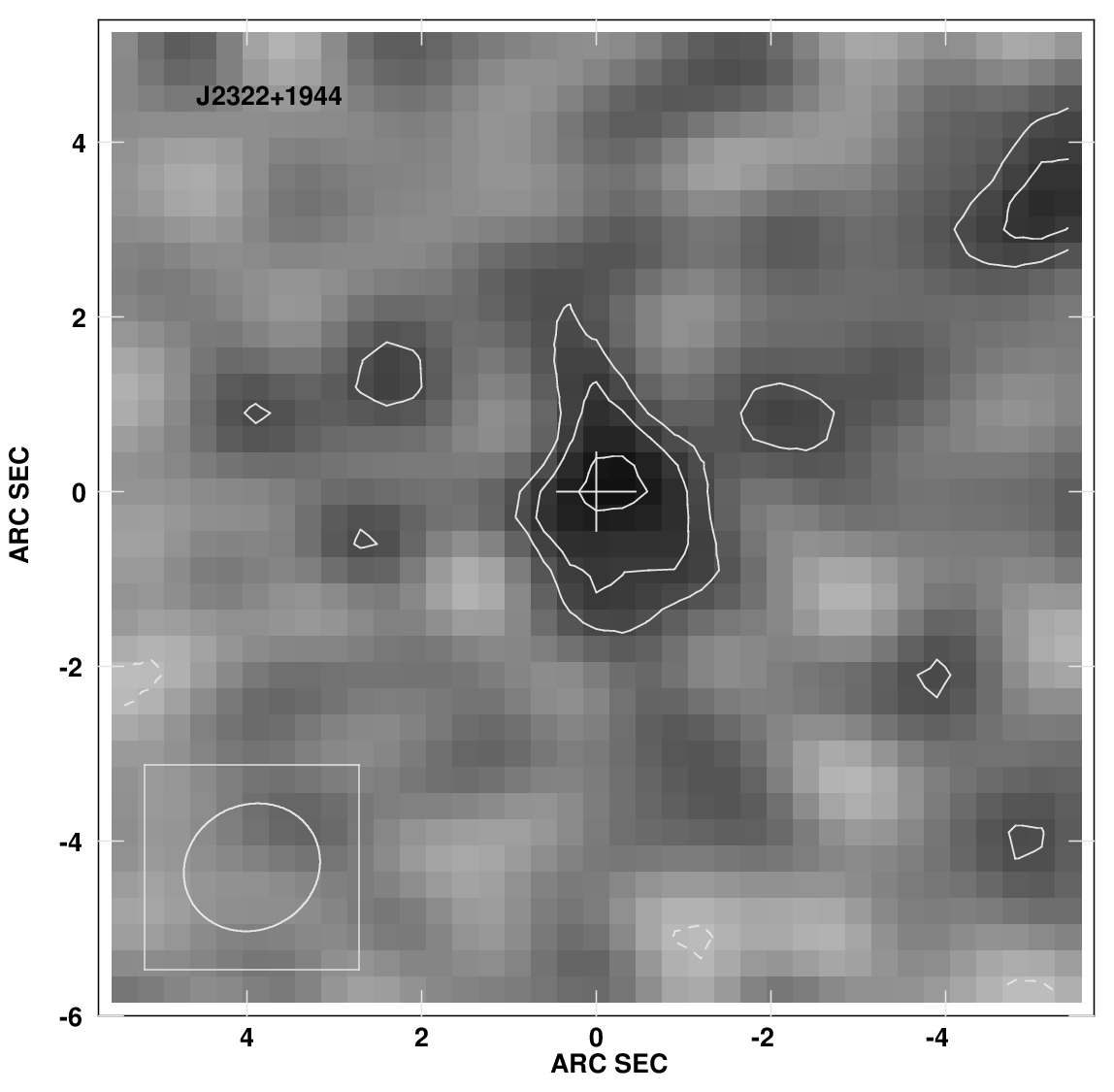,width=4in,angle=-90}
\end{figure}

\clearpage
\newpage

\begin{figure}
\psfig{figure=1048.HIC.PS,width=4in,angle=-90}
\psfig{figure=1057.CHI.PS,width=4in,angle=-90}
\caption{}
\end{figure}

\clearpage
\newpage

\begin{figure}
\psfig{figure=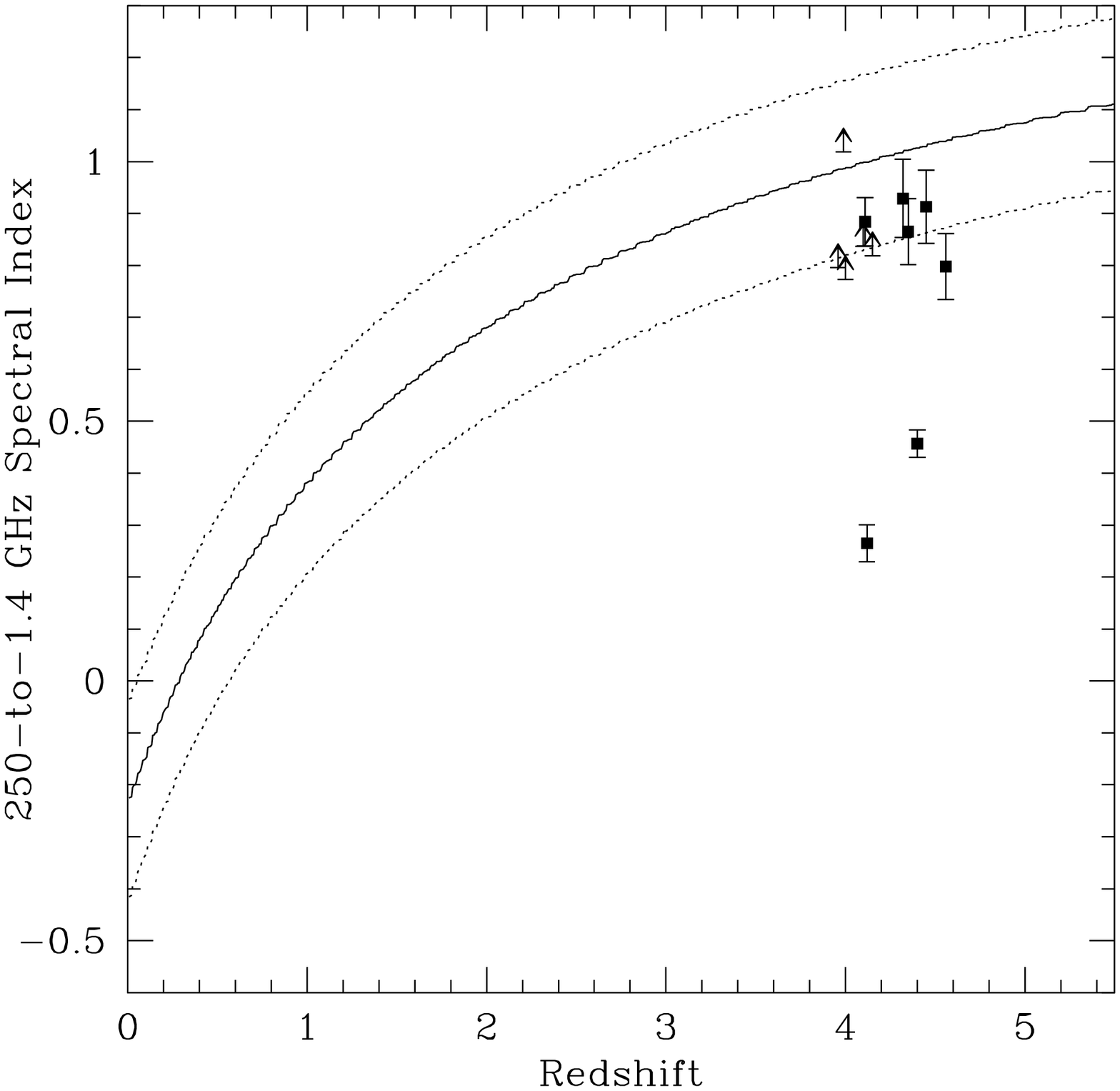,width=5in}
\caption{}
\end{figure}

\clearpage
\newpage

\begin{figure}
\psfig{figure=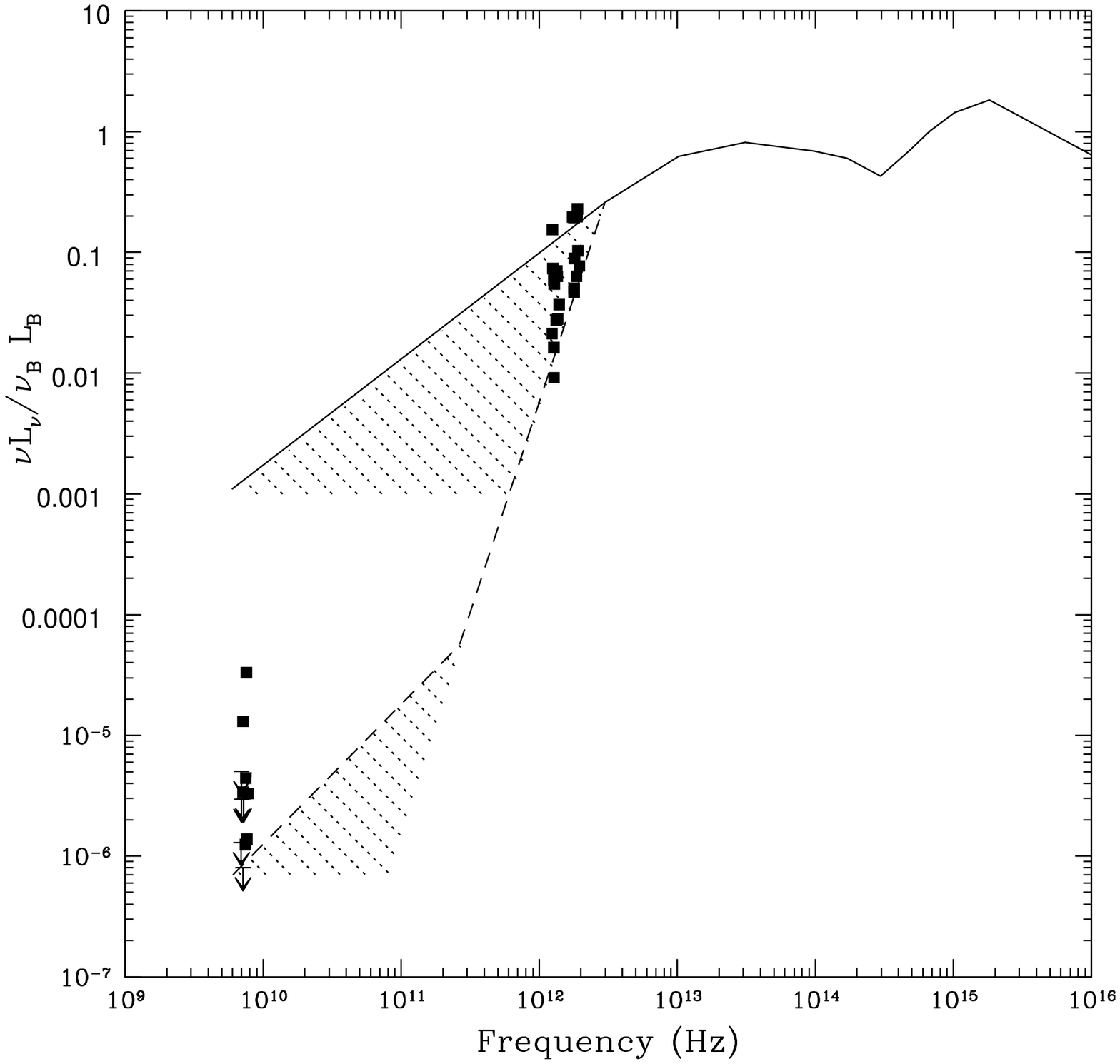,width=5in}
\caption{}
\end{figure}

\clearpage
\newpage

\end{document}